 \documentclass[journal, comsoc]{IEEEtran}

\usepackage{gensymb}
\usepackage{cite}
\usepackage{amsmath,amssymb,amsfonts}
\usepackage{graphicx}
\usepackage{xcolor}
\usepackage{kotex}  
\usepackage{caption}
\usepackage{subcaption}
\usepackage{caption}
\usepackage{subcaption}
\usepackage{tikz}
\usepackage{balance}
\usepackage[font=footnotesize]{caption}
\usepackage{stfloats}
\usepackage{float}
\usepackage{wrapfig}
\usepackage{xcolor}
\usepackage{colortbl} 
\usepackage{color,soul}
\usepackage{url}
\usepackage{verbatim}
\usepackage{graphicx}
\usepackage{cite}

\usepackage{colortbl}
\usepackage{xcolor}
\usepackage{multirow}
\usepackage{array,boldline, makecell,booktabs}
\usepackage{longtable} 

    \usepackage[group-separator={,},group-minimum-digits=4]{siunitx}

\begin{document}
%
\title{CmWave and Sub-THz: Key Radio Enablers and Complementary Spectrum for 6G}

%
%
%

\author{Mayur V. Katwe, Aryan Kaushik, Keshav Singh,  Marco Di Renzo, Shu Sun, Doohwan Lee, Ana G. Armada, Yonina C. Eldar, Octavia A. Dobre, and Theodore S. Rappaport
\thanks{
M. V. Katwe is with the National Institute of Technology, Raipur, India (e-mail: mvkatwe.ece@nitrr.ac.in). \\
$~~~$A. Kaushik is with the School of Engineering \& Informatics, University	of Sussex, UK (e-mail: aryan.kaushik@sussex.ac.uk). \\
$~~~$K. Singh is with the Institute of communications Engineering, National Sun Yat-sen University, Taiwan (e-mail: keshav.singh@mail.nsysu.edu.tw). \\
$~~~$M. Di Renzo is with Universit\'e Paris-Saclay, CNRS, CentraleSup\'elec, France (e-mail: marco.di-renzo@universite-paris-saclay.fr). \\
$~~~$S. Sun is with the Department of Electronic Engineering, Shanghai Jiao Tong University, China (e-mail: shusun@sjtu.edu.cn). \\
$~~~$D. Lee is with the Network Innovation Laboratories, NTT  Corporation, Japan (e-mail: doohwan.lee@ntt.com). 
\\
$~~~$A. G. Armada is with the Department
of Signal Theory and Communications, Universidad Carlos III de Madrid, Spain (e-mail: agarcia@tsc.uc3m.es). \\
$~~~$Y. C. Eldar is with the Faculty of Math and CS, Weizmann Institute of Science, Rehovot, Israel (e-mail: yonina.eldar@weizmann.ac.il).\\
$~~~$O. A. Dobre is with the Faculty of Engineering and Applied Science, Memorial University, Canada (e-mail: odobre@mun.ca). \\
$~~~$T. S. Rappaport is with the Tandon School of Engineering, New York University, USA (e-mail: tlw335@nyu.edu).
}
}

%
%

\markboth{Submitted to IEEE Wireless Communications Magazine}%
{Shell \MakeLowercase{\textit{et al.}}: Bare Demo of IEEEtran.cls for IEEE Journals}

\maketitle

\begin{abstract}
Sixth-generation (6G) networks are poised to revolutionize communication by exploring alternative spectrum options, aiming to capitalize on strengths while mitigating limitations in current fifth-generation (5G) spectrum. This paper explores  the potential opportunities and emerging trends for cmWave and sub-THz spectra as  key radio enablers. {This paper poses and answers three key questions  regarding motivation of additional spectrum to explore the strategic implementation and benefits of cmWave and sub-THz spectra.}  Also, we show using case studies how these complementary spectrum bands will enable new  applications in 6G, such as integrated sensing and communication (ISAC), re-configurable intelligent surfaces (RIS) and non-terrestrial networks (NTN). Numerical simulations reveal that the ISAC performance of cmWave and sub-THz spectra outperforms that of existing 5G spectrum, including sub-6 GHz and mmWave.  Additionally, we illustrate the effective interplay between RIS and NTN to counteract the effects of high attenuation at sub-THz frequencies. Finally, ongoing standardization endeavors, challenges and promising directions are elucidated for these complementary spectrum bands.
\end{abstract}


\IEEEpeerreviewmaketitle

\vspace{-3mm}

\section{Introduction}
\label{sec:1}
\begin{figure}[t!]
\centerline{\includegraphics[width=\linewidth]{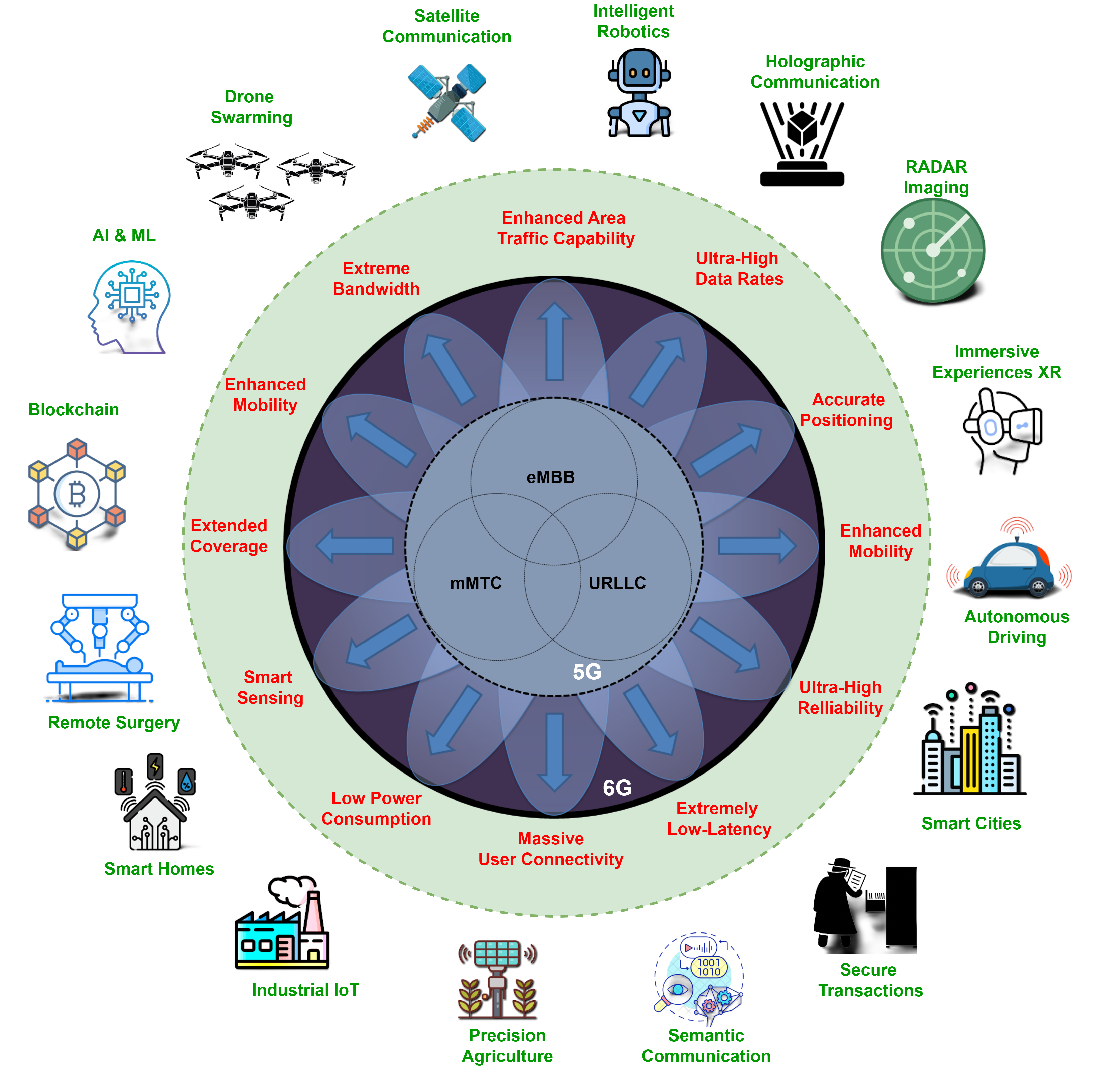}}
	\caption{6G requirements, vision, and applications from WRC-2023, also see \cite{kaushik2024mag,rappaport2019wireless,singh2024towards}.}  
	\label{fig1}
	\vspace{-0.45cm}
\end{figure}

With 5G rolling out across sub-6 GHz and millimeter wave (mmWave) frequencies, sixth-generation (6G) networks  will necessitate and unlock new mobile spectrum bands globally  to accommodate and escalate International Mobile Telecommunications (IMT) 2030 requirements \cite{kaushik2024mag,singh2024towards}. 
6G aims to support massive user densities for humans and devices, potentially millions of devices per square kilometer. Individual throughputs shall exceed 1 Gigabit per second. Additionally, achieving near-instantaneous communication with exceptional network reliability (99.99999$\%$, or five 9's, uptime) is a key goal. In the foreseeable future,  more spectrum will need to be unlocked to   serve  the expansion of a myriad of applications, as shown in Fig. \ref{fig1} \cite{rappaport2019wireless}.

The International Telecommunication Union (ITU) World Radio-communication Conference 2023 (WRC-2023)  convened administrations globally to discuss new spectrum bands and their allocation, including those for 5G Advanced and 6G. For instance, the 7 GHz to 24 GHz band, also called centimeter wave (cmWave) band or ``frequency range 3 (FR3)", is a front-runner (particularly the 7-15 GHz segment) which  offers a compelling combination of broad coverage for widespread connectivity and high data capacity for faster data transfer \cite{shakya2024wideband}. Moreover, 5G millimeter wave (mmWave) spectrum can accommodate slightly higher spectrum deployments such as sub-terahertz (sub-THz) to offer ultra-fast speeds, low-latency, and enhanced user experiences, especially in dense urban environments. Propagation studies, such as \cite{rappaport2019wireless,shakya2024comprehensive,xing2021terahertz,rapapport142}, are crucial for pinpointing bands suitable for mobile and fixed service, a step vital for standardizing enabling technologies and facilitating widespread commercial deployment in a timely manner, while preserving a financially viable ecosystem to attract industry investments.

{ Now, three key questions are posed and answered hereafter to explore the strategic implementation and benefits of cmWave and sub-THz spectra. Addressing these questions will help clarify the motivations, challenges, and advantages inherent in leveraging these frequency bands.}

\textbf{Despite the opportunity for wider IMT identification in the upper 6GHz band and the substantial contiguous spectrum in mmWave, what motivates the push for additional spectrum?}


Reusing existing frequencies for 6G is tempting, but regulations and compatibility issues to achieve IMT 2030 goals make it a complex task. 
While technology allows some temporary sharing, it is simply not economically efficient. With only 700 MHz available in the upper 6 GHz band (6425-7125 MHz), the massive data growth expected for 6G by 2030 may only be supported by new spectrum allocations. Unlike mmWave, which struggles with limited range due to building penetration losses \cite{rappaport2019wireless}, cmWave strikes an excellent balance between capacity and coverage. 
Based on work and public comments filed by the Millimeter Wave Coalition, a group of companies and universities founded in 2018, it is evident that mmWave bands are emerging, especially for fixed wireless access (FWA), backhaul, and high-density venues, and spectrum below 100 GHz is very crowded, often allocated to military or satellite applications, making large contiguous swaths of bandwidth (e.g. 500 MHz or more) extremely rare \cite{shakya2024comprehensive,xing2021terahertz}.
 Having much wider bandwidth allocations to concatenate at sub-THz would offer significantly higher capacity and data rates, while also sharing  similar propagation characteristics with mmWave, making it an attractive option for future wireless communications. 

\textbf{Does network densification necessitate a shift from omnidirectonal antenna and sub-6Ghz spectrum to high-gain directional antenna and high-carrier frequency spectrum?
}


  Indeed, small cells offer much better power efficiency than larger cells when considering energy per bit\cite{rappaport2024waste}. However, the increased deployment of small cells, essential for densification, raises concerns for power consumption and environmental sustainability.  Finding suitable locations for numerous small cells complicates network planning and deployment, and achieving high user experience through extreme densification increases the utility bill and negatively impacts the carbon footprint.
Small cells require specific separation distances to prevent interference and ensure seamless user experiences during handovers, which can be facilitated by tighter adaptive beam technology in mmWave and sub-THz frequencies.  Countries like  USA, Japan, and others are rapidly adopting  mmWave based fixed wireless access, with over 1 billion users, replacing traditional fiber and copper infrastructure. Nevertheless, increasing base station density at higher frequencies improves communication efficiency, but benefits plateau beyond a certain point\cite{rappaport2024waste}. Directional antennas mitigate free space losses in higher-frequency networks but escalate power consumption with more base stations. To boost network efficiency without sacrificing throughput, transitioning to higher carrier frequencies with high-gain antennas is recommended \cite{rappaport2024waste}.





\textbf{Why cmWave spectrum should be complemented with sub-THz spectrum and not with mmWave and higher THz frquencies?}

The aggregate available  spectrum is  immensely greater at the sub-THz and terahertz (THz) range \cite{xing2021terahertz}, yet cmWave spectrum, when augmented with sub-THz spectrum is advantageous because sub-THz offers both outdoor, outdoor-to-indoor penetration, and indoor and short-range urban deployments, unlike the very short range of higher THz frequencies alone. Sub-THz  provides superior sensing and communication capabilities that will become part of all cellphone devices in the coming decade \cite{rappaport2019wireless}. Additionally, cmWave can leverage existing infrastructure given similar propagation properties to sub-6 GHz deployments \cite{shakya2024comprehensive}, thereby reducing deployment costs and providing better coverage and penetration, making it a practical and cost-effective choice for widespread 6G roll-out. For instance, cmWave can be used for fronthaul and sub-THz can be preferred for backhual links between cell sites, as well as for mobile access (e.g., integrated access and backhaul, a new feature in the 5G  standard).

{ In this article, we present the opportunities and emerging trends of cmWave and sub-THz spectra as key radio enablers for 6G. We detail the complementary spectrum and discuss its key usage scenarios, providing case studies on new applications like integrated sensing and communication (ISAC), reconfigurable intelligent surfaces (RIS), and non-terrestrial networks (NTN). Additionally, we discuss ongoing industry experimentation, standardization efforts, future directions, and challenges for these spectrum bands.
\begin{figure*}[t!]
\centerline{\includegraphics[width=\linewidth]{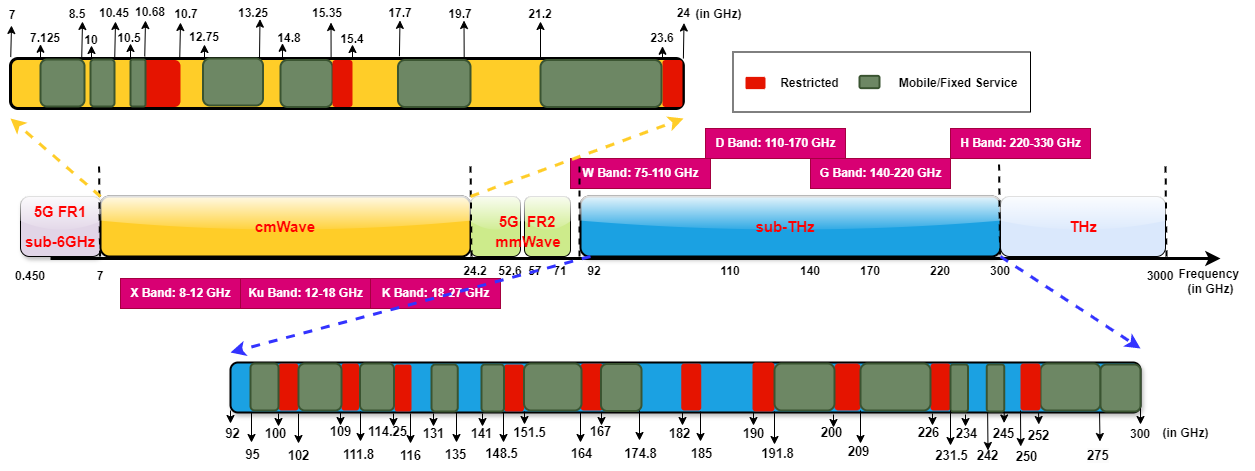}}
	\caption{CmWave and sub-THz Spectrum.}
	\label{fig2}
\end{figure*}
\section{Potential and Ongoing Efforts in CmWave and Sub-THz}
As per third-generation partnership project (3GPP) Release (Rel.) 17, 5G FR2 has been allocated up to 71 GHz, spanning from 24.25 GHz to 71 GHz. This frequency region encompasses the 60 GHz unlicensed band, known as the E band (60 GHz to 90 GHz). As it can be observed from Fig. \ref{fig2}, the  blue bands represent sub-THz frequency regions, i.e. 92-300 GHz,  identified for potential future wireless communications by the ITU WRC-2019 and WRC-2023.
\subsection{Lower cmWave band (7-15 GHz)}
WRC-2023 Resolution 256 along with WRC-2027 agenda Item 1.2 highlights potential IMT identification for bands such as 6.425 – 7.025 GHz in Region 1 (Europe), 7.025 – 7.125 GHz globally, and 10 – 10.5 GHz in Region 2 (North America). NATO bands encompass 7.25 – 8.4 GHz, 8.5 – 10.5 GHz, 13.4 – 14 GHz, 14.62 – 15.23 GHz, 15.7 – 17.7 GHz, and 20.2 – 21.2 GHz, among others. Fixed  wireless links currently occupy various bands, including 6.4 – 7.1 GHz, 7.4 – 7.9 GHz, 7.9 – 8.5 GHz, and 12.75 – 13.25 GHz, with potential clearance possibilities available through further fiber rollout to replace the incumbent fixed wireless links in certain markets. Recognizing this shift, the U.S. FCC recently designated a portion of the 12.2-13.25 GHz band (specifically 12.7-13.25 GHz) for mobile services, while also seeking input on sharing the remaining segment (12.2-12.7 GHz) for more robust terrestrial operations. It is important to note that  regulatory efforts dealing with the removal of or sharing with incumbents, including satellite and fixed wireless license holders are  crucial in identifying new spectrum for IMT in the extended mid-band 7-15 GHz range, including 7.125-8.5 GHz, 12.75-13.25 GHz, and 14.8-15.35 GHz, to support the cellular mobile industry.   

\begin{figure*}[t!]
\centerline{\includegraphics[width=\linewidth]{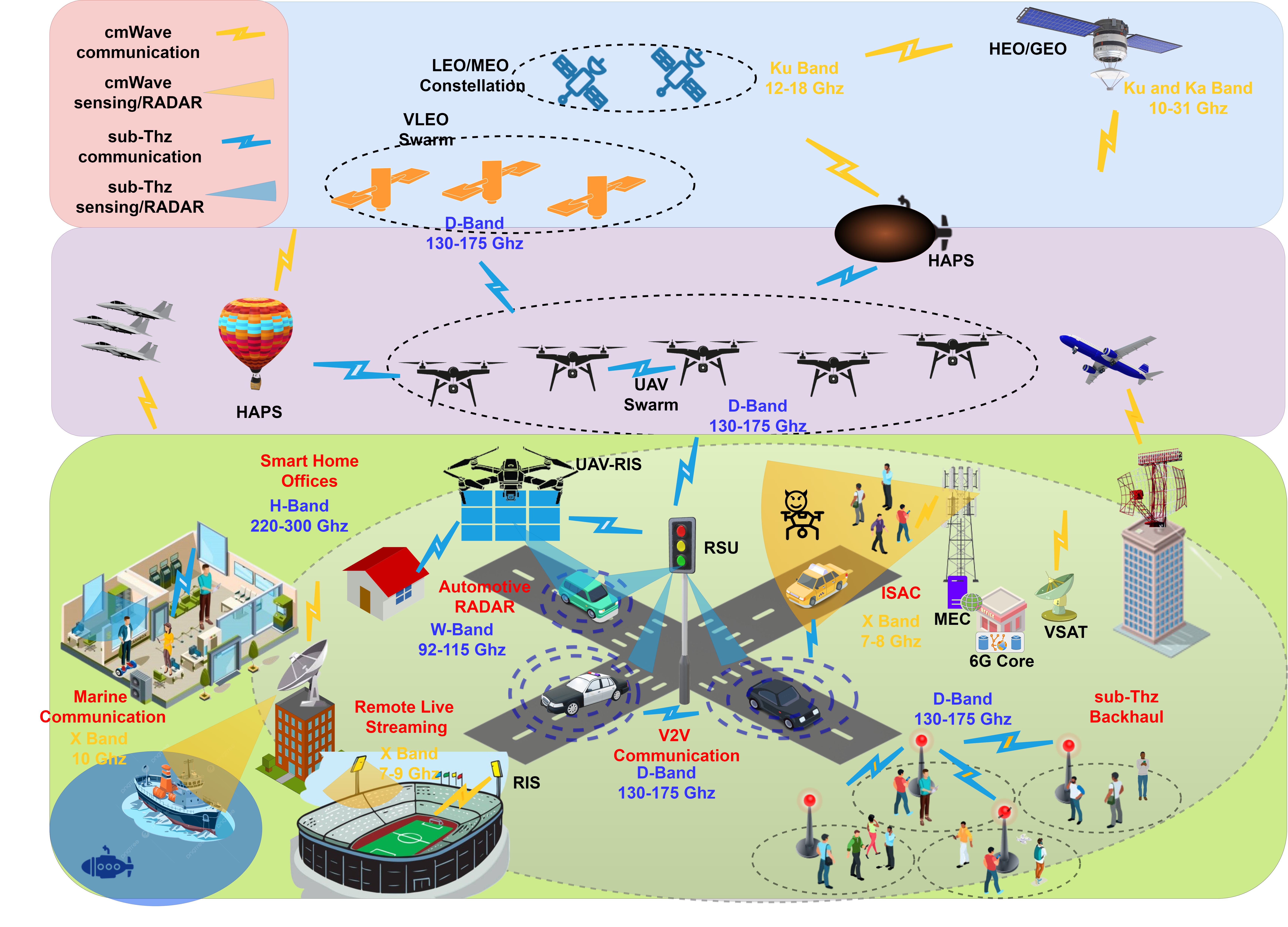}}
	\caption{Potential usage scenarios for cmWave and sub-THz spectrum for mobile access.}
	\label{fig_app}
\end{figure*}

\subsection{Sub-THz Spectrum (92-300 GHz)}
Prospective frequency bands within the sub-THz spectrum, including the D band (110 GHz to 170 GHz), G band (140 GHz to 220 GHz), and the H/J band (220 GHz to 330 GHz), have been identified during WRC-2023. Resolutions 255,663 and 721 in WRC-2023 outline the necessity of specific spectrum ranges, such as 102-109.5 GHz, 151.5-164 GHz, 167-174.8 GHz, 209-226 GHz, and 252-275 GHz, for the advancement of the next generation of mobile communications \cite{xing2021terahertz}. Additionally, the spectrum from 275-296 GHz and the terahertz spectrum are designated without specific conditions to protect Earth exploration-satellite service (passive) applications. It is anticipated that a combination of licensed (including W and D bands) and unlicensed bands within the sub-THz range will meet the diverse requirements of 6G use cases and deployment types \cite{ericsson20246G}. Presently, the development of terahertz-range bands for wireless backhaul/access is underway, with focus on the W band spectrum (75 GHz to 110 GHz) and the D band (110 GHz to 170 GHz), particularly by base station infrastructure vendors. Research efforts for 6G are shifting towards exploring the D band and the H/J band around 300 GHz.

\section{CmWave and sub-THz: Applications and Synergies with 6G Usage Scenarios}
\label{sec:4}
Undoubtedly, the cmWave and sub-THz  spectrum bands offer unique opportunities for a variety of advanced applications and key 6G usage scenarios. Fig. \ref{fig_app} showcases the potential  interplay of cmWave and sub-THz spectrum which  paves the way for deploying highly interconnected mobile networks that offer high data rates, exceptional reliability, and ultra-low latency.

\subsection{CmWave and sub-THz for Integrated Sensing and Communications(ISAC)}
{ ISAC is an emerging application of cmWave and sub-THz spectrum, merging wireless communication with high-precision sensing. 
Both positioning and sensing benefit from the large bandwidths and antenna arrays available in the cmWave and sub-THz bands. With a per-operator contiguous bandwidth of 200 MHz in the cmWave band, sensing and time-of-flight measurements can achieve a range resolution of less than a meter, whereas mmWave and sub-THz carriers permit ranging to within a mm or so\cite{rappaport2019wireless}. } Seamless physical, digital, and virtual interactions are elevating human augmentation to new heights through pervasive, low-power integrated communication and sensing \cite{marco_survey_THz}. This high-accuracy positioning and sensing are crucial for applications like the metaverse, digital twins, and collaborative robots (cobots), driving the need for reliability in complex industrial environments.  However, key questions remain, such as how to sense the environment using communication systems without degrading performance, and how to design signals and share resources between communication and sensing. 

\subsection{Automotive RADAR and V2X Communication}
The dynamic nature of transportation traffic and the ever-increasing data bandwidth demands pose significant challenges for achieving high transmission rates in vehicle-to-everything (V2X) networks. These networks require higher data rates with very low latency, and modulations that are robust to fast channel variations.  Most current commercial automotive radars operate within the  industrial, scientific and medical (ISM) 77 GHz band, and more recently, the 76-81 GHz band. However, in the automotive context, the current resolution and information provided by mm-wave radar may not suffice for many applications. Therefore, exploring the sub-THz band for automotive radar and V2X networks presents a significant opportunity for improvement in resolution and detail due to its correspondence with an atmospheric absorption window (e.g., see \cite{rappaport2019wireless} and \cite{xing2021terahertz}). 
Additionally, the Rayleigh criterion dictates that at such high frequencies, many common surfaces appear rough to low-THz radar, leading to diffuse surface scattering. This allows reflections from the same radar target area to be intercepted from various positions with different aspect angles, making sub-THz radar more sensitive to surface textures of road objects and enabling finer resolution radar images.  Sub-THz bands can deliver speeds over 100 Gbps within 50 m, ideal for V2X data transfers at traffic lights. 

\subsection{X-haul Networks}
Another technical motivation is the need for wireless backhaul/fronthaul that can provide very high throughput, catering to both mobile and fixed or nomadic scenarios. Together, the three radio access network (RAN) links$—$backhaul (core and the central unit), midhaul (between distributed units), and fronthaul (between central and distributed unit)  is referred to as X-haul. A dense deployment of X-haul cannot rely solely on optical fiber; it requires the flexibility and cost-efficiency provided by wireless links, which are economically viable and faster to deploy since they avoid the civil works and roadway disruptions that can delay projects in protected areas, like historical quarters. The anticipated high data traffic necessitates a capacity of tens of Gb/s. Achieving this data rate wirelessly requires very wide frequency channels in previously untapped bands at the mmWave spectrum and above, such as the E-band (10 GHz over 71-76 GHz and 81-86 GHz) and the sub-THz spectrum (over 100 GHz in the range of 110-310 GHz). 
Although E-band front ends with a capacity of a few Gb/s are currently available, higher throughput is needed for fiber-like performance. Significant efforts are underway globally to develop equipment and front ends at sub-THz frequencies capable of tens of Gb/s. 


\subsection{Non-Terrestrial Networks}
The cmWave spectrum has traditionally been utilized extensively for satellite services or NTN. Currently, low earth orbit (LEO)   satellites operate in Ku-band, Ka-band, and Q-/V-bands. The interplay of mega LEO constellations with high-altitude platform systems (HAPS) and unmanned aerial vehicles (UAVs)  are expected to play a crucial role in the upcoming 6G networks.  To meet future high data rate demands, NTN will need to utilize higher frequencies in the W-band and THz frequencies above 300 GHz. At these frequencies, antennas can achieve high directional gain with narrow beams, supporting high-capacity systems and secure communications with reduced risk of eavesdropping. 
 However, 
atmospheric gases, primarily water vapor and oxygen, cause significant frequency-selective molecular absorption losses at sub-THz frequencies. This absorption is more pronounced compared to lower frequency bands like the G-band. To counteract these high absorption and path losses, highly directional antennas can be deployed, which benefit from low beam divergence and smaller size for the same gain, allowing for the use of massive antenna arrays and can  maintain effective communication. 
Molecular absorption by atmospheric gases also introduces noise by re-radiating absorbed power back into the communication channel. 
Moreover, the Earth-space link is power-limited, requiring efficient use of transmitted power from LEO satellites, which may be facilitated by modulations that are robust to non-linearities. 
\subsection{Reconfigurable Intelligent Surfaces (RIS)-aided Communication}
Reconfigurable intelligent surface (RIS)-aided  communication, essentially modern-day repeaters or re-transmitters, may be incorporated for a wide range of frequencies, and at cmWave  and sub-THz spectra offer complementary advantages. The cmWave provides more robust coverage and penetration, ideal for ensuring reliable connectivity in diverse environments, including areas with obstacles and extensive urban settings. 
Meanwhile, sub-THz offers ultra-high data rates and improved sensing capabilities, enabling high-capacity, short-range communication and precise environmental awareness. The number of RIS elements for a given fixed area can be increased by shifting from sub-6 GHz to cmWave, and from mmWave to sub-THz. Many experimental trials have shown that for sub-THz frequencies, the reconfigurable elements on RIS are physically very small, allowing for the incorporation of large-scale RIS even with smaller sizes. Additionally, RIS can help resolve attenuation issues with sub-THz frequencies by intelligently directing and reflecting signals, thus extending the  effective coverage   range \cite{wan2021terahertz}.

\begin{figure}[t!]
\begin{subfigure}{\linewidth}
 \centering
		\includegraphics[scale=0.30]{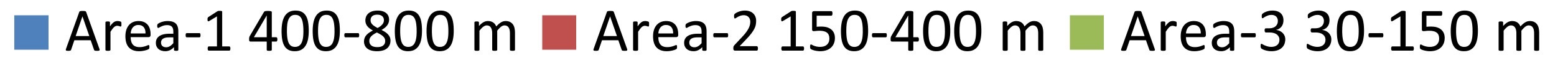}
	\end{subfigure}
  \begin{subfigure}{.495\linewidth}
 \centering
		\includegraphics[scale=0.13]{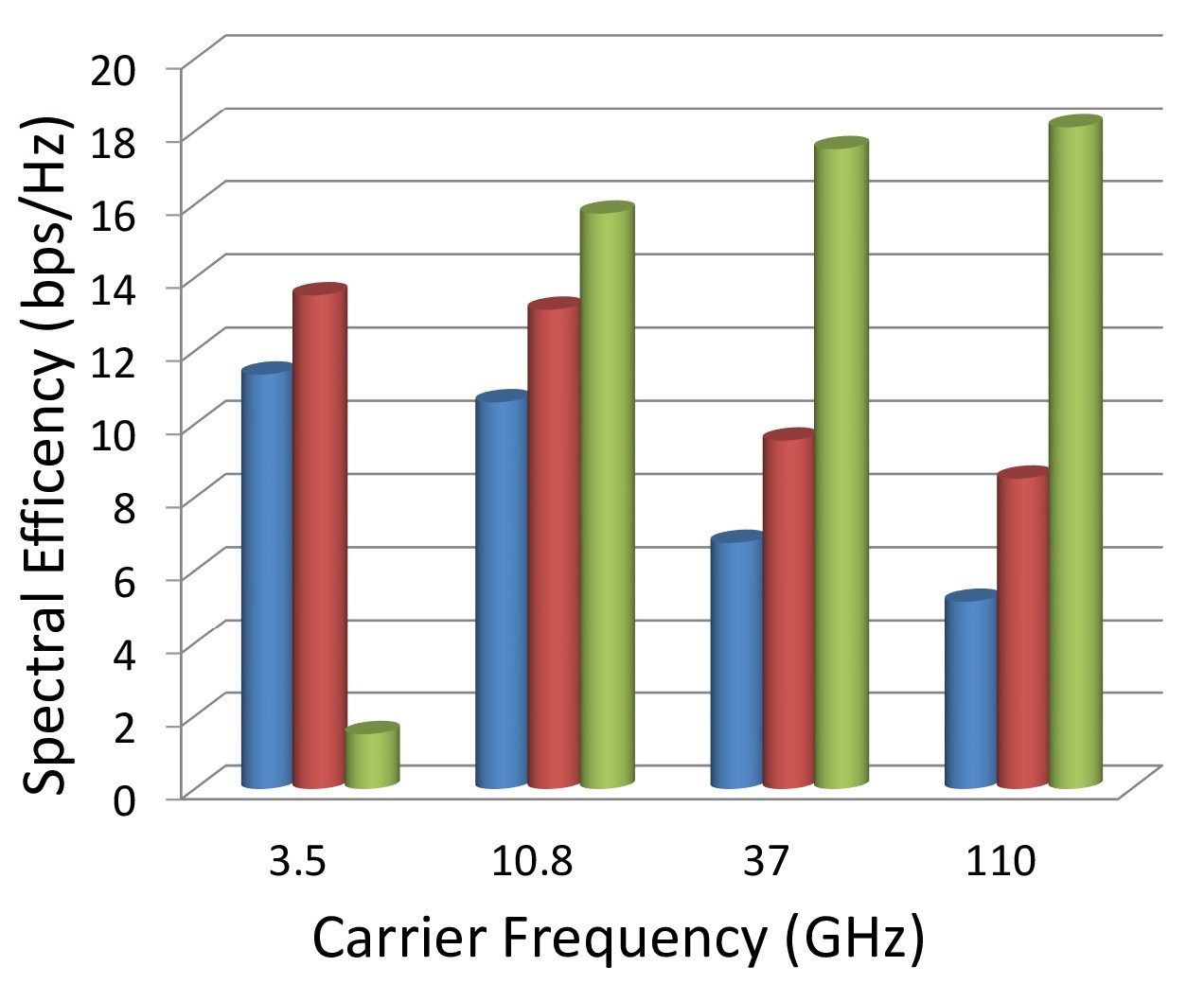}
	\end{subfigure}
  \begin{subfigure}{.495\linewidth}
 \centering
		\includegraphics[scale=0.13]{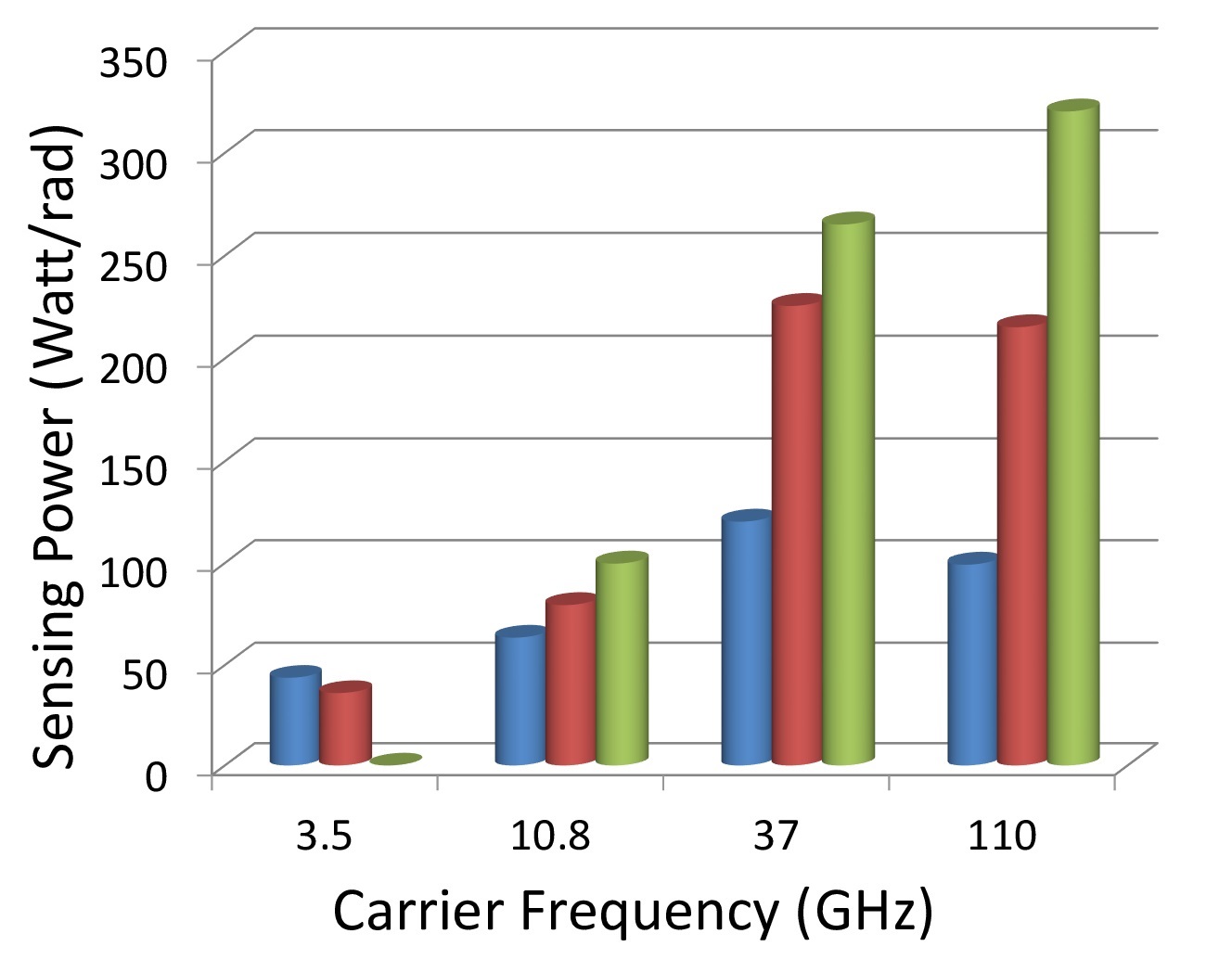}
		
	\end{subfigure}
 
 \caption{{ISAC performance for various carrier frequencies when  2 targets  and  3 single-antenna  users are in line-of-sight with uniform-linear array  antenna at 0.25 m with ISAC model and adaptive beamforming design as described in \cite{wang2022noma}.}}
 \label{isac_fig}
\end{figure}
\section{Case Studies}
Next, we  describe case studies of this complementary spectrum which offer insights into the diverse applications and their prospects for 6G networks.
\subsection{ISAC with cmWave and sub-THz}
Here, we analyze the performance of ISAC  framework across various spectra, encompassing sub-6 GHz, cmWave, mmWave, and sub-THz, operating within distinct geographical deployment (areas), as shown in Fig. \ref{isac_fig}. 
{The ISAC performance is assessed under a varying free-space path loss model for different carrier frequencies with a fixed size of antenna array. Evidently, the spectral efficiency per unit bandwidth diminishes with an increase in carrier frequency due to increased attenuation. Nevertheless, the increase in carrier frequency allows to accommodate higher antennas which ameliorate communication and sensing performance.
Likewise, the sub-THz spectrum offers superior sensing capabilities compared to mmWave, particularly in short ranges (Area-3).}
However, for mid-range applications (Area-2 of Fig. \ref{isac_fig}), mmWave may be preferable over sub-THz. In comparison to sub-6 GHz, the cmWave spectrum provides enhanced sensing capabilities while maintaining similar communication capabilities. This improvement is particularly notable for short and mid-range applications, where the sub-6 GHz band may encounter challenges such as low-resolution beamforming and inter-user interference mitigation.
\begin{figure}[t!]
\centerline{\includegraphics[width=\linewidth]{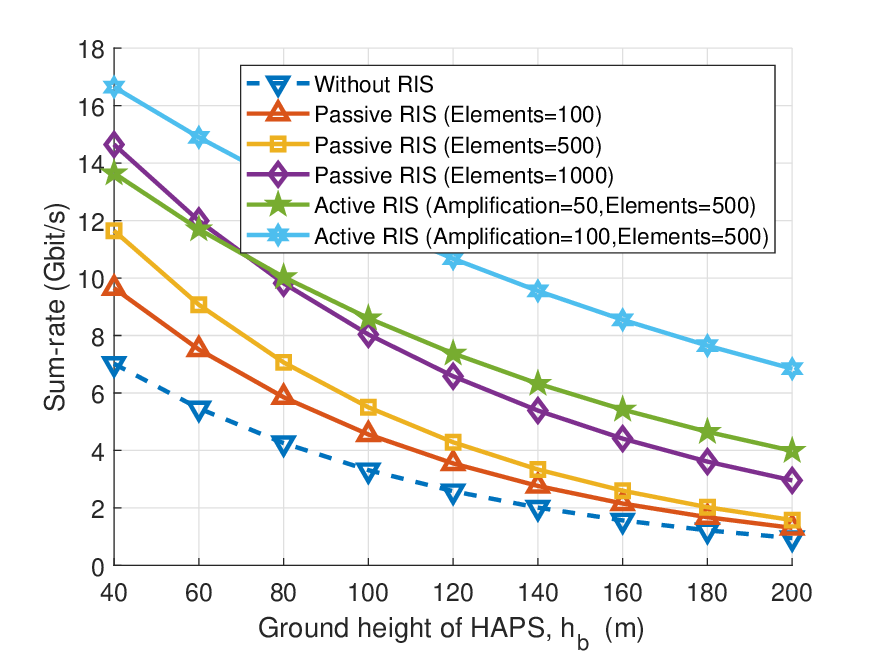}}
	\caption{Performance evaluation of RIS-assisted  NTN at 142 GHz with $h_u=30$ m, 4 UAVs, 3 HAPS, 7 BSs and path loss model as in  \cite{xing2021terahertz}.  BSs, UAVs and HAPS are randomly deployed in the $600\times 600$ m$^2$ deployment region with UAVs and HAPS are hovering at fixed height of $h_u$ and $h_b$ (in m), respectively.}  
	\label{ntn_fig}
\end{figure}
\subsection{RIS-aided Cell-free NTN with sub-THz Communication}
{To validate the future potential of NTN for sub-THz, we analyze the impact of the sub-THz spectrum using numerical simulation. In a given a fully-connected  cell-free 64$\times$64 multiple-input multiple-output (MIMO) NTN, a set of HAPSs   performs  downlink communication with a set of ground base-stations (BSs)  via a swarm of UAV-integrated RIS.   } Fig. \ref{ntn_fig} presents a performance evaluation of the average sum-rate  of the system with varying the height of HAPS. As expected, the spectral performance significantly degrades with increasing the distance  between BS and HAPSs due to the strong  attenuation of the sub-THz signal.
Furthermore, the HAPS-integrated passive RIS is effective only within a range of 80-90 m, where its performance is comparable to the direct link scenario (without RIS). To achieve better performance than direct links, a higher number of RIS elements is recommended. Additionally, active RIS, sometimes also considered similar to holographic surfaces, can provide superior performance even with a lower number of RIS elements, as shown in Fig. \ref{ntn_fig}.  

\subsection{Airy Beams: Full Utilization of Electromagnetic Waves with sub-THz}
Shortened wavelengths in the sub-THz range extend the near-field range to hundreds of meters, allowing for advanced beam control that leverages electromagnetic near-field phenomena such as orbital angular momentum, bent propagation, and non-diffractive propagation \cite{lee2022multishape}. The manipulation of sub-THz electromagnetic waves using RIS enables more customized beam control and the application of Airy and Bessel beams, which have not been extensively utilized until now.
\begin{figure}[t!]
\centerline{\includegraphics[width=0.9\linewidth]{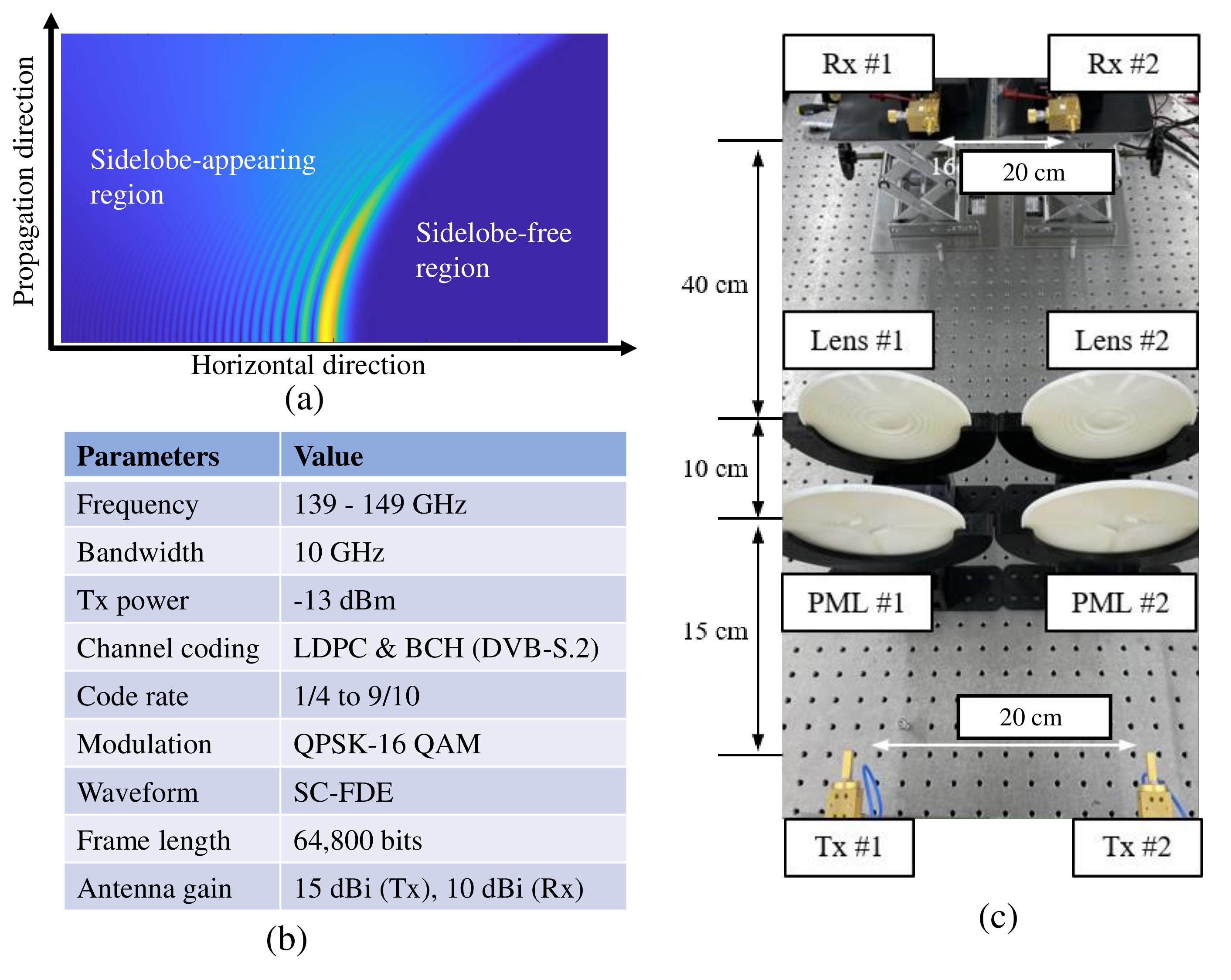}}
	\caption{(a) Propagation of the Airy wave and its sidelobe characteristics, (b) Experimental parameters, (c) Experimental setup of parallel wireless transmission using Airy beam (PML: phase modulation lens).}  
	\label{airybeam_fig}
\end{figure}
Fig. \ref{airybeam_fig} shows the unique properties of Airy beams with asymmetric sidelobe distributions and sidelobe-free regions, the experimental parameters at sub-THz frequencies, and the experimental setup for two-stream transmission using Airy beams, respectively. The total transmission rate for two-stream transmission without using Airy beams was 25.33 Gbit/s due to interference. In contrast, the total transmission rate for two-stream transmission using Airy beams was 46.67 Gbit/s, by exploiting sidelobe-free regions. We have experimentally confirmed that the Airy beam enables parallel transmission of multiple streams without traditional MIMO channel estimation and equalization. 
\section{Industry Perspectives, Standardization and Experimentation on CmWave/sub-THz}
Securing sufficient 6G spectrum requires collaboration among vendors, mobile network operators (MNO), regulators, service representatives, and research organizations to assess needs, recommend regulatory changes, and balance societal requirements.
\subsection{Field Trials and Industry Efforts on sub-THz}
The authors in \cite{rapapport142} carried out two extensive outdoor wideband measurement campaigns in downtown Brooklyn, focusing on the sub-THz band at 140 GHz with transmitter-receiver separation distances up to 117.4 m. These campaigns included: i) a terrestrial urban microcell measurement campaign, and ii) a rooftop surrogate satellite and backhaul measurement campaign. Japan has fostered pioneering work at sub-THz frequencies.
NTT DOCOMO, INC., NTT Corporation, NEC Corporation, and Fujitsu Limited have collaboratively announced the creation of an advanced wireless device capable of achieving ultra-high-speed transmissions of 100 Gbps in the 100 GHz and 300 GHz bands. DOCOMO has developed wireless transmission equipment that can deliver data rates of up to 100 Gbps over a distance of 100 m.
Other telecom vendors such as Ericsson, Nokia, Keysight, and Qualcomm have developed  a testbed RAN system operating in the sub-THz frequency range, specifically in the 92–100 GHz band, at D-Band (142 GHz) and H/J-Band (285 GHz), which is capable of achieving peak throughputs exceeding 100 Gbps \cite{ntt_exp}.

 \subsection{Standardization Aspects of cmWave and sub-THz bands}
A study item was made available in Release 16 for 3GPP in Jan 2021 \cite{3gpp_cmwave}, which discusses the potential for cmWave for 5G spectrum.  IEEE Std. 802.15.3d marks a significant milestone as the inaugural IEEE-family standard tailored for wireless communications spanning up to 69-GHz-wide channels within the sub-THz spectrum range, specifically from 253 to 322 GHz. While spectrum access can be secured through ITU WRC decisions, regional agreements, or country-specific allocations, harmonizing frequency bands and technical specifications globally or regionally is essential for economies of scale and benefits to consumers and businesses. 
Although an IMT identification does not impose an obligation for implementation in any particular country, it serves as a pivotal step in harmonizing IMT frequency bands, signaling to the Information and Communications Technology (ICT) industry the need for equipment development. For instance, the designation of the frequency range from 14.62 to 15.23 GHz as a harmonized NATO band for fixed and mobile services underscores the importance of coordinated efforts through WRC resolutions and regional decisions. Moreover, the ITU's framework establishes a structured process for evaluating and establishing technical conditions for various frequency bands worldwide, incorporating sharing studies to mitigate harmful interference between different services. 
Thus, the ITU WRCs remain the preferred avenue for comprehensive harmonization. However, given the time-intensive nature of the process, with an IMT identification preceding regional licensing, efforts to finalize ITU's work must commence well before 2030. This urgency underscores the ongoing research endeavors in both industry and academia, aiming to validate the usability and technical feasibility of cmWave and sub-THz spectrum by WRC-2027.

\section{Future Prospects and Considerations}
\label{sec:5}
Some future prospects and considerations for cmWave and sub-THz technologies are as follows:
\begin{itemize}
    \item \textbf{Spectrum sharing and regulation:} A significant portion of the cmWave spectrum is already allocated to incumbents for various applications, such as satellite communications and radar. Implementing 6G will require careful planning and coexistence mechanisms to avoid interference with these existing services. 
    International agreements and national regulations must be established to ensure efficient spectrum utilization. Also, regulations need to permit higher effective isotropic radiated power (EIRP) limits. This would compensate for increased propagation losses in these bands and allow operators to reuse existing 4G and 5G tower sites, which is essential for the cost-effective roll-out of new technology.
    \item \textbf{Multi-functional RIS}: RISs are vital for sub-THz communications but need smart artificial intelligence (AI)-assisted controllers and algorithms. Future RIS should switch between modes and integrate sensors to train AI controllers for a wide range of mmWave and sub-THz bands. Sensors could be part of the RIS hardware or operate in a lower frequency band as network controlled repeaters. As discussed in Fig. \ref{ntn_fig}, active RISs with power amplifiers can boost sub-THz links and improve system performance by amplifying signals before reflection.
    \item\textbf{Devices  for mass-market THz wireless}: The sub-THz spectrum presents unique hurdles for semiconductor components, requiring the creation of new high-frequency, high-power output components suitable for mass adoption. This includes advancements in power amplifiers and other critical components that influence system performance, such as output power, efficiency, and bandwidth. 
    \item \textbf{Channel sounding and modelling}: Understanding electromagnetic wave propagation in sub-THz and THz frequencies is crucial for robust communication system development. Research on channel propagation above 100 GHz is vital due to the impact of human bodies, vehicles, and environmental conditions like rain. Refining existing 5G channel models to incorporate these factors is necessary, emphasizing the importance of innovative THz measurement instruments in 6G research.
    \item \textbf{Modulation and channel estimation}: Robust modulations that can work under high channel variability and non-linear amplification are required for the use cases that we have detailed. Also needed are ways to obtain the required channel estimation for coherent demodulation and for ISAC purposes, and the design of the corresponding reference signals. Whether it is best to have them organized and processed in the time, frequency or delay-Doppler domain strongly depends on the use case.
\end{itemize}

\section{Conclusion}
\label{sec:6}
The article has presented a comprehensive analysis of the potential role of cmWave and sub-THz spectra as key enablers for 6G wireless communication networks. Through discussions on their applications as well as emerging spectrum allocation trends, we have underscored the role of both low band cmWave and high band mmWave and sub-THz for the future of wireless communication. Case studies on ISAC, as well as NTNs, have illustrated the vast potential for new services and capabilities, both on a network level and in a consumer device of the future. Numerical simulations given here have demonstrated how the combination of cmWave and THz spectrum offers  superior capabilities  compared to current 5G spectra. Importantly, these studies are crucial for pinpointing bands suitable for mobile and fixed services, a vital step  for standardizing enabling technologies and facilitating widespread commercial deployment in a timely manner.
\bibliographystyle{IEEEtran}
\bibliography{IEEEabrv,BibRef}

\begin{thebibliography}{10}
\providecommand{\url}[1]{#1}
\csname url@samestyle\endcsname
\providecommand{\newblock}{\relax}
\providecommand{\bibinfo}[2]{#2}
\providecommand{\BIBentrySTDinterwordspacing}{\spaceskip=0pt\relax}
\providecommand{\BIBentryALTinterwordstretchfactor}{4}
\providecommand{\BIBentryALTinterwordspacing}{\spaceskip=\fontdimen2\font plus
\BIBentryALTinterwordstretchfactor\fontdimen3\font minus
  \fontdimen4\font\relax}
\providecommand{\BIBforeignlanguage}[2]{{%
\expandafter\ifx\csname l@#1\endcsname\relax
\typeout{** WARNING: IEEEtran.bst: No hyphenation pattern has been}%
\typeout{** loaded for the language `#1'. Using the pattern for}%
\typeout{** the default language instead.}%
\else
\language=\csname l@#1\endcsname
\fi
#2}}
\providecommand{\BIBdecl}{\relax}
\BIBdecl

\bibitem{kaushik2024mag}
A.~Kaushik, R.~Singh, S.~Dayarathna, R.~Senanayake, M.~Di~Renzo, M.~Dajer,
  H.~Ji, Y.~Kim, V.~Sciancalepore, A.~Zappone, and W.~Shin, ``Toward integrated
  sensing and communications for {6G}: {Key} enabling technologies,
  standardization, and challenges,'' \emph{IEEE Commun. Stds. Mag.}, vol.~8,
  no.~2, pp. 52--59, June 2024.

\bibitem{singh2024towards}
R.~Singh, A.~Kaushik, W.~Shin, M.~Di~Renzo, V.~Sciancalepore, D.~Lee,
  H.~Sasaki, A.~Shojaeifard, and O.~A. Dobre, ``Towards {6G} evolution: {Three}
  enhancements, three innovations, and three major challenges,'' \emph{arXiv
  preprint arXiv:2402.10781}, Feb. 2024.

\bibitem{rappaport2019wireless}
T.~S. Rappaport, Y.~Xing, O.~Kanhere, S.~Ju, A.~Madanayake, S.~Mandal,
  A.~Alkhateeb, and G.~C. Trichopoulos, ``Wireless communications and
  applications above 100 {GHz}: {Opportunities} and challenges for {6G} and
  beyond,'' \emph{IEEE Access (Invited Paper)}, vol.~7, pp. 78\,729--78\,757,
  June 2019.

\bibitem{shakya2024wideband}
\BIBentryALTinterwordspacing
{D. Shakya, M. Ying, T. S. Rappaport, Hitesh Poddar, Peijie Ma, Yanbo Wang, and
  Idris Al-Wazani}, ``Propagation measurements and channel models in indoor
  environment at {6.75 GHz FR1(C)} and {16.95 GHz FR3} upper-mid band spectrum
  for {5G and 6G},'' in \emph{IEEE Global Communications Conference (GLOBECOM),
  Cape Town, South Africa}.\hskip 1em plus 0.5em minus 0.4em\relax IEEE, Dec
  2024, pp. 1--6. [Online]. Available: \url{https://arxiv.org/pdf/2405.01358v1}
\BIBentrySTDinterwordspacing

\bibitem{shakya2024comprehensive}
D.~Shakya, M.~Ying, T.~S. Rappaport, H.~Poddar, P.~Ma, Y.~Wang, and
  I.~Al-Wazani, ``Comprehensive {FR3} and {FR1 (C)} upper-mid band propagation
  and material penetration loss measurements and channel models in indoor
  environment for {5G} and {6G},'' \emph{TecharXiv preprint}, pp. 1--22, 2024.

\bibitem{xing2021terahertz}
Y.~Xing and T.~S. Rappaport, ``Terahertz wireless communications: {Co}-sharing
  for terrestrial and satellite systems above 100 {GHz},'' \emph{IEEE Commun.
  Lett.}, vol.~25, no.~10, pp. 3156--3160, 2021.

\bibitem{rapapport142}
S.~Ju and T.~S. Rappaport, ``142 {GHz} multipath propagation measurements and
  path loss channel modeling in factory buildings,'' in \emph{IEEE Int. Conf.
  Commun.}, 2023, pp. 5048--5053.

\bibitem{rappaport2024waste}
T.~S. Rappaport, M.~Ying, N.~Piovesan, A.~De~Domenico, and D.~Shakya, ``Waste
  factor and waste figure: {A} unified theory for modeling and analyzing wasted
  power in radio access networks for improved sustainability,'' \emph{arXiv
  preprint arXiv:2405.07710}, May 2024.

\bibitem{ericsson20246G}
E.~Semaan, E.~Tejedor, R.~K. Kochhar, S.~Magnusson, and S.~Parkvall, ``6g
  spectrum- enabling the future mobile life beyond 2030,'' Ericsson, Tech.
  Rep., May 2024.

\bibitem{marco_survey_THz}
W.~Jiang, Q.~Zhou, J.~He, M.~A. Habibi, S.~Melnyk, M.~El-Absi, B.~Han, M.~D.
  Renzo, H.~D. Schotten, F.-L. Luo, T.~S. El-Bawab, M.~Juntti, M.~Debbah, and
  V.~C.~M. Leung, ``Terahertz communications and sensing for {6G} and beyond:
  {A} comprehensive review,'' \emph{IEEE Commun. Surv. \& Tuts., Early Access
  Article}, pp. 1--31, 2024.

\bibitem{wan2021terahertz}
Z.~Wan, Z.~Gao, F.~Gao, M.~Di~Renzo, and M.-S. Alouini, ``Terahertz massive
  {MIMO} with holographic reconfigurable intelligent surfaces,'' \emph{IEEE
  Trans. Commun.}, vol.~69, no.~7, pp. 4732--4750, 2021.

\bibitem{wang2022noma}
Z.~Wang, Y.~Liu, X.~Mu, Z.~Ding, and O.~A. Dobre, ``{NOMA} empowered integrated
  sensing and communication,'' \emph{IEEE Commun. Lett.}, vol.~26, no.~3, pp.
  677--681, 2022.

\bibitem{lee2022multishape}
D.~Lee, Y.~Yagi, and H.~Shiba, ``Multishape radio: {New} approach to utilizing
  the physical properties of electromagnetic waves,'' \emph{IEICE
  Communications Express}, vol.~11, no.~9, pp. 571--576, 2022.

\bibitem{ntt_exp}
\BIBentryALTinterwordspacing
{NTT Release}. (April 2024) {DOCOMO, NTT, NEC and Fujitsu} develop top-level
  sub-terahertz {6G} device capable of ultra-high-speed 100 {Gbps}
  transmission. [Online]. Available:
  \url{https://group.ntt/en/newsrelease/2024/04/11/240411a.html}
\BIBentrySTDinterwordspacing

\bibitem{3gpp_cmwave}
\BIBentryALTinterwordspacing
{3GPP Release 16}, ``Study on the 7 to 24 {GHz} frequency range for {NR},''
  Tech. Rep. (TR) 38.820, V16.1.0., 2021. [Online]. Available:
  \url{https://portal.3gpp.org/desktopmodules/Specifications/SpecificationDetails.aspx?specificationId=3599}
\BIBentrySTDinterwordspacing

\end{thebibliography}


\end{document}